\documentclass[twocolumn,prc,showpacs,preprintnumbers,superscriptaddress]{revtex4-1}

\usepackage{amsmath} 
\usepackage{amssymb}
\usepackage{amsfonts}
\usepackage{mathrsfs}
\usepackage{units}
\usepackage{multirow}

\usepackage{dcolumn}
\usepackage{bm}
\usepackage{rotating} 

\makeatletter
\def\overbracket#1{\mathop{\vbox{\ialign{##\crcr\noalign{\kern3\p@}
    \downbracketfill\crcr\noalign{\kern3\p@\nointerlineskip}
    $\hfil\displaystyle{#1}\hfil$\crcr}}}\limits}
\def\downbracketfill{$\m@th
    \makesm@sh{\llap{\vrule\@height.7\p@\@depth2.3\p@\@width.7\p@}}%
    \leaders\vrule\@height.7\p@\hfill
    \makesm@sh{\rlap{\vrule\@height.7\p@\@depth2.3\p@\@width.7\p@}}$}
\makeatother
\newcommand{\disregard}[1]{}

\newcommand{\hata}{\hat{a}}
\renewcommand{\hata}{{a}}

\begin{document}

\title{Simple regularization scheme for multi-reference density functional theories}

\author{W. Satu{\l}a}
\affiliation{Institute of Theoretical Physics, Faculty of Physics, University of Warsaw, ul. Ho\.za
69, PL-00-681 Warsaw, Poland}
\affiliation{Helsinki Institute of Physics, P.O. Box 64, FI-00014 University of Helsinki, Finland}

\author{J. Dobaczewski}
\affiliation{Institute of Theoretical Physics, Faculty of Physics, University of Warsaw, ul. Ho\.za
69, PL-00-681 Warsaw, Poland}
\affiliation{Helsinki Institute of Physics, P.O. Box 64, FI-00014 University of Helsinki, Finland}
\affiliation{Department of Physics, P.O. Box 35 (YFL),
University of Jyv\"askyl\"a, FI-40014  Jyv\"askyl\"a, Finland}

\date{\today}

\begin{abstract}
\begin{description}
\item[Background]
Extensions of single-reference (SR) energy-density-functionals (EDFs)
to multi-reference (MR) applications involve using the generalized
Wick theorem (GWT), which leads to singular energy kernels that
cannot be properly integrated to restore symmetries, unless the EDFs
are generated by {\it true} interactions.

\item[Purpose]
We propose a new method to regularize the MR EDFs, which is based on
using auxiliary quantities obtained by multiplying the kernels with
appropriate powers of overlaps.

\item[Methods]
Regularized matrix elements of two-body interactions are obtained by
integrating the auxiliary quantities and then solving simple linear
equations.

\item[Results]
We implement the new regularization method within the self-consistent
Skyrme-Hartree-Fock approach and we perform a proof-of-principle
angular-momentum projection (AMP) of states in odd-odd nucleus
$^{26}$Al. We show that for EDFs generated by {\it true}
interactions, our regularization method gives results identical to
those obtained within the standard AMP procedure. We also show that
for EDFs that do not correspond to {\it true} interactions, it gives
stable and converging results that are different than unstable and
non-converging standard AMP values.

\item[Conclusions]
The new regularization method proposed in this work may provide us
with a relatively inexpensive and efficient tool to generalize SR
EDFs to MR applications, thus allowing for symmetry restoration and
configuration mixing performed for typical nuclear EDFs, which most
often do not correspond to {\it true} interactions.

\end{description}
\end{abstract}

\maketitle

\section{Introduction}\label{intro}

Density Functional Theory (DFT) is a universal approach used in
quantum chemistry, molecular physics, and condensed matter physics to
calculate properties of electronic systems. Its extension to nuclear
physics is by no means trivial, encountering difficulties associated
in part with the binary composition of atomic nuclei, spin-dependent
interactions, superfluidity, and strong surface effects. Major
difference between the electronic and nuclear DFT is associated with
the lack of external binding potential, as atomic nuclei are
self-bound systems, and with the saturation of nuclear interactions
at a given value of density. This implies that the nuclear DFT must
necessarily be formulated in terms of intrinsic, and not laboratory
densities, which, in turn, leads to the spontaneous breaking of
fundamental symmetries.

In spite of these difficulties, the nuclear DFT is the microscopic
tool of choice to study in a systematic manner medium-mass and heavy
nuclei. In the symmetry-broken mean-field variant, often referred to
as single-reference (SR) DFT, the method has proven to be extremely
successful in reproducing and predicting bulk nuclear properties like
masses, quadrupole moments, or nuclear radii. However, for a precise
description of numerous observables, the SR nuclear DFT is inadequate.
In particular, at the SR level, matrix elements of electromagnetic
transitions or beta decays can only be treated within a
quasiclassical approximation. Fully quantal calculations of such
observables are impossible without the symmetry restoration, which
requires extensions from the SR to multi-reference (MR) DFT.

However, within the MR DFT, implementation of the symmetry
restoration is plagued with technical and conceptual difficulties
\cite{(Dob07),(Lac09),(Ben09)}. The reason is that the SR DFT,
serving as the starting point, is usually derived from an effective,
density-dependent pseudo-potential, and is therefore not directly
related to a Hamiltonian. The only reasonable, and to a large extent
unambiguous generalization of the SR energy density functional (EDF)
to the MR level is possible within the generalized Wick's theorem
(GWT)~\cite{(Bla86)} that establishes a one-to-one
correspondence between the SR and MR functionals. In such an
implementation, the MR EDF retains the form of the underlying SR EDF,
but is solely expressed in terms of the so-called transition
densities. Unfortunately, the resulting MR EDFs are, in general,
singular and require regularization. In spite of preliminary
attempts, concentrating mostly on a direct removal of self-pairing
effects~\cite{(Lac09),(Ben09)}, the problem of regularization still
lacks satisfactory and practical solution. The aim of this work is
to propose such a solution.

The paper is organized as follows. In Sec.~\ref{MRDFT}, we recall the
standard formulation of the MR DFT scheme based on the GWT, and we
identify sources of potential pathologies. Then, in
subsections~\ref{LRS} and~\ref{QRS} we present two variants of the
new regularization scheme, hereafter called linear (LR) and quadratic
regularization (QR), respectively. Our method is illustrated by
applications to the angular-momentum projection (AMP) problem, but
can similarly be used to restore other broken symmetries. In
particular, the particle-number restoration within the
pairing-plus-quadrupole Hamiltonian is currently studied in
Ref.~\cite{(Raf14)}. Summary and perspectives are discussed in
Sect.~\ref{sum}.

\section{Regularization scheme}\label{RS}

\subsection{Standard MR DFT scheme\/}\label{MRDFT}

In this work, all implementations of projection methods are based on the GWT,
which allows for deriving compact and numerically tractable expressions for off-diagonal
matrix elements between Slater determinants. For an arbitrary Slater determinant $|\Psi\rangle$,
by $|\tilde{\Psi}\rangle = \hat R(\Omega )|\Psi\rangle$ we
denote the one that is rotated in space, gauge space, or isospace, for the angular-momentum,
particle-number, or isospin restoration, respectively. Hereafter, we focus our attention
on the AMP, but the presented ideas and methodology can be rather
straightforwardly generalized to the particle-number~\cite{(Raf14)} or isospin projections.

In order to bring forward the origin of singularities in energy kernels~\cite{(Dob07),(Lac09),(Ben09)},
it is instructive to recall principal properties
of the standard GWT approach. Let us start with a one-body density-independent operator
$\hat F = \sum_{ij} F_{ij} \hata_i^\dagger \hata_j $. Its off-diagonal kernel (the matrix element
divided by the overlap), can be calculated
with the aid of GWT, and reads~\cite{(Bla86)}:
\begin{eqnarray}\label{eqn:1}
   \label{1B}
   \frac{ \langle\Psi | {\hat{F}}|\tilde{\Psi}\rangle}{\langle\Psi |\tilde{\Psi}\rangle}
           =  \sum_{ij}\; F_{ij} \overbracket{\hata^+_i \hata_j} \equiv
              \sum_{ij}\; F_{ij} \tilde{\rho}_{ji},
\end{eqnarray}
where
\begin{equation}\label{rhotr}
   \tilde{\rho}_{ji} \equiv
   \overbracket{a^+_{i} a_{j}} \equiv \frac{\langle\Psi | a^+_{i}
   a_{j}|\tilde{\Psi}\rangle}
                                               {\langle\Psi|\tilde{\Psi}\rangle},
\end{equation}
denotes transition density matrix. Therefore,
its matrix element between the unprojected state $|\Psi\rangle$ and AMP state
$|IMK\rangle = \hat{P}^I_{MK} | \Psi\rangle$ can be calculated from
\begin{eqnarray}\label{eqn:3}
 F_{IMK} & \equiv &  \langle\Psi | {\hat{F}} \hat{P}^I_{MK} | \Psi\rangle = \nonumber \\
  & = & \frac{2I+1}{8\pi^2} \int d\Omega\, D^{I\, ^\star}_{MK} (\Omega )
      \langle\Psi | {\hat{F}}|\tilde{\Psi}\rangle,
\end{eqnarray}
where
\begin{equation}\label{projop}
\hat{P}^I_{MK}  =  \frac{2I+1}{8\pi^2 } \int
 D^{I\, *}_{M K}(\Omega ) \hat{R}(\Omega ) \, d\Omega
\end{equation}
is the AMP operator, $D^{I}_{M K}(\Omega )$
is the Wigner function, and $\hat R (\Omega ) = e^{-i\alpha \hat I_z} e^{-i\beta \hat I_y} e^{-i\gamma \hat I_z}$
stands for the active rotation operator in space,
parametrized in terms of Euler angles $\Omega = (\alpha, \beta, \gamma )$,
and $M$ and $K$ denote the angular-mo\-men\-tum components along the
laboratory and intrinsic $z$-axis, respectively~\cite{(RS80),(Var88)}.

The immediate conclusion stemming from
Eqs.~(\ref{eqn:1})--(\ref{rhotr}) is that the overlaps, which appear
in the denominators of the matrix element and transition density matrix,
cancel out, and the matrix element $\langle\Psi |
{\hat{F}}|\tilde{\Psi}\rangle$ of an arbitrary one-body
density-independent operator $\hat F$ is free from singularities and
can be safely integrated, as in Eq.~(\ref{eqn:3}).

Let us now turn our attention to two-body operators.
The most popular two-body effective interactions used in nuclear structure calculations are the zero-range
Skyrme~\cite{(Sky56),(Sky59)} and finite-range Gogny~\cite{(Gog75)} effective forces. Because of their explicit density dependence,
they should be regarded, for consistency reasons,  as generators of two-body part of the nuclear
EDF. The transition matrix element of the two-body generator reads:
\begin{equation}
   \label{V2B-1}
  \langle\Psi|\hat V_{{\rm 2B}}|\tilde{\Psi}\rangle
       = \frac{1}{4}\sum_{ijkl} \bar{V}_{ijkl}\left[\tilde{\rho}\right]\;
         \langle\Psi|\hata^+_i \hata^+_j \hata_l \hata_k|\tilde{\Psi}\rangle,
\end{equation}
where $\bar{V}_{ijkl}\left[\tilde{\rho}\right]$  denotes the antisymmetrized
transition-density-dependent matrix element.
Gogny and Skyrme effective interactions both
contain local terms proportional to $ \rho^{\,\eta} $ which, in the
MR DFT formulation, are usually replaced with the transition (mixed)
density $ \rho^{\,\eta} \rightarrow \tilde \rho^{\,\eta}$~\cite{(Bon90)}. Such a
procedure, although somewhat arbitrary, is very common, because it
fulfills a set of internal consistency criteria  formulated in
Refs.~\cite{(Rob07),(Rob10)}. These include hermiticity, independence
of scalar observables on the orientation of the intrinsic system, and
consistency with the underlying mean field. The alternative way of
proceeding is to substitute density-dependent terms with projected
density~\cite{(Sch04)} or average density~\cite{(Dug03)}. These
scenarios do not fulfill the consistency criteria and will not be
discussed here.

Evaluating the transition matrix element, Eq.~(\ref{V2B-1}), with the aid of GWT, one obtains,
\begin{eqnarray}
   \label{V2B-2}
   \frac{\langle\Psi|\hat V_{{\rm 2B}}|\tilde{\Psi}\rangle}{\langle\Psi|\tilde{\Psi}\rangle}
      & = & \frac{1}{4}\sum_{ijkl} \bar{V}_{ijkl}\left[\tilde{\rho}\right]\;
         \left(   \overbracket{\hata^+_i \hata^+_j}\;\overbracket{\hata_l \hata_k}\; \right.
         \nonumber \\
             & + & \left. \;\overbracket{\hata^+_i \hata_k}\; \overbracket{\hata^+_j \hata_l}\;
              - \;\overbracket{\hata^+_i \hata_l}\; \overbracket{\hata^+_j \hata_k}\; \right).
\end{eqnarray}
Furthermore, for particle-number-conserving theory,
contractions $\overbracket{a^+_i a^+_j}$ and
$\overbracket{a_l a_k}$ vanish, whereas the remaining two contractions give
products of two transition density matrices,
\begin{eqnarray}
   \label{V2B-2a}
   \frac{\langle\Psi|\hat V_{{\rm 2B}}|\tilde{\Psi}\rangle}{\langle\Psi|\tilde{\Psi}\rangle}
      & = & \frac{1}{4}\sum_{ijkl} \bar{V}_{ijkl}\left[\tilde{\rho}\right]\;
         \left(   \tilde{\rho}_{ki}\tilde{\rho}_{lj}
              -   \tilde{\rho}_{li}\tilde{\rho}_{kj} \right),
\end{eqnarray}
or
\begin{eqnarray}
   \label{V2B-3}
   \frac{\langle\Psi|\hat V_{{\rm 2B}}|\tilde{\Psi}\rangle}{\langle\Psi|\tilde{\Psi}\rangle}
      & = & \frac{1}{4}\sum_{ijkl} \bar{V}_{ijkl}\left[\tilde{\rho}\right]\;
         \left( \frac{\langle\Psi|\hata^+_i \hata_k|\tilde{\Psi}\rangle\;
                      \langle\Psi|\hata^+_j \hata_l|\tilde{\Psi}\rangle}{\langle\Psi|\tilde{\Psi}\rangle^2}
       \right. \nonumber \\
      & - &  \left.   \frac{\langle\Psi|\hata^+_i \hata_l|\tilde{\Psi}\rangle\;
                      \langle\Psi|\hata^+_j \hata_k|\tilde{\Psi}\rangle}{\langle\Psi|\tilde{\Psi}\rangle^2}
         \right),
\end{eqnarray}
that is, the transition matrix element reads
\begin{eqnarray}
   \label{V2B-4}
   \langle\Psi|\hat V_{{\rm 2B}}|\tilde{\Psi}\rangle
      & = & \frac{1}{2}\sum_{ijkl} \bar{V}_{ijkl}\left[\tilde{\rho}\right]\;
         \frac{\langle\Psi|\hata^+_i \hata_k|\tilde{\Psi}\rangle\;
               \langle\Psi|\hata^+_j \hata_l|\tilde{\Psi}\rangle}{\langle\Psi|\tilde{\Psi}\rangle} .
\end{eqnarray}
This defines the matrix element between the unprojected and AMP states,
\begin{eqnarray}\label{vimk}
 V^{{\rm 2B}}_{IMK} & \equiv & \langle\Psi | {\hat{V}_{{\rm 2B}}} \hat{P}^I_{MK} | \Psi\rangle = \nonumber \\
  & = & \frac{2I+1}{8\pi^2} \int d\Omega\, D^{I\, ^\star}_{MK} (\Omega )
      \langle\Psi | {\hat{V}_{{\rm 2B}}}|\tilde{\Psi}\rangle .
\end{eqnarray}
At variance with the one-body case discussed above, the integrand in
Eq.~(\ref{vimk}) is inversely proportional to the overlap, thus
containing potentially dangerous (singular) terms. The singularity
disappears only if the sums in the numerator, evaluated at angles
$\Omega$ where the overlap $\langle\Psi|\tilde{\Psi}\rangle$ equals
zero, give a vanishing result; such a cancellation requires
evaluating the numerator without any approximations or omitted terms.
An additional singularity is created by the density dependence of the interaction.

If some approximation of the numerator is involved, the leading-order singularity goes as:
\begin{equation}
\langle\Psi | {\hat{V}_{{\rm 2B}}}|\tilde{\Psi}\rangle
 \sim \frac{1}{ \langle\Psi | \tilde{\Psi}\rangle ^{1+\eta}} ,
\end{equation}
with the term $1/6\leq \eta \leq 1$ inherited from the direct density
dependence of the Gogny or Skyrme effective forces which are, as
already mentioned, commonly used to generate the modern
non-relativistic nuclear EDFs. This singularity precludes, in
general, determination of the integral in  Eq.~(\ref{vimk}). Only in
special cases, e.g., for signature-symmetry conserving states in
even-even nuclei~\cite{(Zdu07),(Vaq13)}, the overlaps never vanish
and the problem does not appear.

Thus, the GWT formulation of MR DFT is, in general, singular. In
fact, it is well defined only for ${\hat{V}_{{\rm 2B}}}$ being a {\it
true\/} interaction. An example of such an EDF generator is the
density-independent Skyrme interaction SV$_{\rm T}$, which is the SV
interaction of Ref.~\cite{(Bei75)} with all the EDF tensor terms
included (these were omitted in the original definition of SV).
Interaction SV$_{\rm T}$ was recently used to calculate the
isospin-symmetry-breaking corrections to superallowed $0^+\rightarrow
0^+$ $\beta$-decay by means of the isospin- and angular-momentum
projected DFT formalism~\cite{(Sat11)}.

Progress in development of projection techniques and difficulties in
working out reliable regularization schemes for density-dependent
interactions~\cite{(Lac09)} increased the demand for
density-independent effective interactions and stimulated vivid
activity in this field resulting in developing density-independent
zero-range~\cite{(Was12),(Sad13)} as well as
finite-range~\cite{(Ben13a)} forces. The spectroscopic quality of
these new forces is, however, still far from satisfactory. In
addition, the technology of performing beyond-mean-field calculations
with these novel interactions is being developed only
now~\cite{(Bal14),(Bal14a)}.

The pathologies arising in the GWT description of the two-body energy kernels come from
uncompensated zeros of the overlap matrix. The central idea of this work is to cure the
problem by replacing the calculation of projected matrix elements with higher-order
quantities, which are regularized by multiplying the integrands with an appropriately chosen power
of the overlap:
\begin{equation}
\langle \Psi |\hat V_{\rm{2B}} | \tilde{\Psi}\rangle \rightarrow \langle \Psi |\hat V_{\rm{2B}} | \tilde{\Psi}\rangle
\langle \Psi | \tilde{\Psi}\rangle^n .
\end{equation}

The proposed regularization scheme amounts to
replacing the calculation of matrix elements $V^{{\rm 2B}}_{IMK}$, given in Eq.~(\ref{vimk}), by the calculation of an auxiliary
quantities defined as:
\begin{equation}\label{etilde}
V^{{\rm 2B,n}}_{IMK} = \frac{2I+1}{8\pi^2} \int d\Omega\, D^{I\, ^\star}_{MK} (\Omega )
\langle \Psi |\hat V_{\rm{2B}} | \tilde{\Psi}\rangle \langle \Psi | \tilde{\Psi}\rangle^n .
\end{equation}
The central assumption behind such a regularization method is that the
two-body matrix element $\langle \Psi |\hat V_{\rm{2B}} |
\tilde{\Psi}\rangle $ is regularizable, meaning that there exists a
regularization procedure allowing for removal of singularities and
replacing the infected matrix elements by regular ones,
\begin{equation}\label{reg}
\langle \Psi |\hat V_{\rm{2B}} | \tilde{\Psi}\rangle \longrightarrow
\widetilde{ \langle \Psi |\hat V_{\rm{2B}} | \tilde{\Psi}\rangle } ,
\end{equation}
for which the projected matrix elements can be calculated as in Eq.~(\ref{vimk}),
\begin{eqnarray}\label{timk}
\tilde{V}^{{\rm 2B}}_{IMK} & = & \frac{2I+1}{8\pi^2} \int d\Omega\, D^{I\, ^\star}_{MK} (\Omega )
      \widetilde{\langle\Psi | {\hat{V}_{{\rm 2B}}}|\tilde{\Psi}\rangle} ,
\end{eqnarray}
and which, in turn, can be expanded on a series of the Wigner $D$-functions:
\begin{equation}\label{skex}
\widetilde{ \langle \Psi |\hat V_{\rm{2B}} | \tilde{\Psi}\rangle } = \sum_{I'M'K'} \tilde{V}_{I'M'K'}^{\rm{2B}}
D^{I'}_{M' K'} (\Omega ).
\end{equation}
Indeed, by inserting expansion (\ref{skex}) into Eq.~(\ref{timk}),
and employing the orthonormality conditions of the Wigner
$D-$functions~\cite{(Var88)}, one straightforwardly obtains the
desired result.

Finally, the regularized matrix elements $\widetilde{ \langle \Psi
|\hat V_{\rm{2B}} | \tilde{\Psi}\rangle }$ are determined be
requesting that the auxiliary quantities (\ref{etilde}), calculated
before and after regularization are equal, that is,
\begin{equation}\label{true}
V^{{\rm 2B,n}}_{IMK} \equiv \tilde{V}_{IMK}^{\rm{2B,n}}.
\end{equation}
for
\begin{equation}\label{ttilde}
\tilde{V}^{{\rm 2B,n}}_{IMK} = \frac{2I+1}{8\pi^2} \int d\Omega\, D^{I\, ^\star}_{MK} (\Omega )
\widetilde{ \langle \Psi |\hat V_{\rm{2B}} | \tilde{\Psi}\rangle} \langle \Psi | \tilde{\Psi}\rangle^n .
\end{equation}
Let
us underline that our method does not require any explicit {\it a
priori\/} knowledge of the regularization scheme. Also note that the
expansion coefficients $\tilde{V}^{{\rm 2B}}_{IMK}$, appearing in
Eqs.~(\ref{timk}) and~(\ref{skex}), represent {\it true\/}
regularized (two-body) matrix elements.

Two variants of such a regularization scheme, dubbed linear
regularization (LR) and quadratic regularization (QR), corresponding
to, respectively, $n=1$ and $n=2$, are discussed below in
Secs.~\ref{LRS} and~\ref{QRS}.

\subsection{Linear regularization scheme}\label{LRS}

The LR scheme applies, for example, to the density-independent SV
interaction in its original formulation~\cite{(Bei75)}, that is,
without the EDF tensor terms. It is also applicable to the
density-dependent SIII functional~\cite{(Bei75)}. The reason is that
the density-dependence of this latter force does not lead, for a
given type of particles, to a higher power of density. The third
example, where the LR should be sufficient, is the Coulomb exchange treated
in the so called Slater approximation~\cite{(Sla51)}, because in this
approximation the exchange Coulomb transition matrix element behaves as
\begin{equation}
  \tilde{\rho}^{\, 4/3} \, \langle \Psi | \tilde{\Psi}\rangle \sim \langle \Psi | \tilde{\Psi}\rangle^{-1/3}.
\end{equation}
In cases when the LR scheme is sufficient to cancel all poles of the
integrand, the set of auxiliary quantities $V^{{\rm 2B,1}}_{IMK}$
(\ref{etilde}) can be calculated, in principle exactly, using suitably
chosen quadratures with sufficient number of integration nodes.

The overlap is always regular and, therefore, it can be expanded in terms of the Wigner $D$-functions as,
\begin{equation}\label{ovex}
\langle \Psi | \tilde{\Psi}\rangle = \sum_{I''M''K''} c_{I''M''K''}^{\rm{{\cal N}}}
D^{I''}_{M'' K''} (\Omega ),
\end{equation}
where
\begin{equation}
c_{IMK}^{\rm{{\cal N}}} \equiv \langle \Psi | \hat{P}^I_{MK} | \Psi \rangle =
\frac{2I+1}{8\pi^2} \int d\Omega\, D^{I\, ^\star}_{MK} (\Omega ) \langle \Psi | \tilde{\Psi}\rangle .
\end{equation}
Substitution of expansions (\ref{skex}) and (\ref{ovex}) into Eq.~(\ref{ttilde}), taken at $n=1$, leads to:
\begin{eqnarray}\label{etilde2}
\tilde{V}^{{\rm 2B,1}}_{IMK} = \frac{2I+1}{8\pi^2} \sum_{I'M'K'} \tilde{V}_{I'M'K'}^{\rm{2B}}
\sum_{I''M''K''} c_{I''M''K''}^{\rm{{\cal N}}} \nonumber \\
\int d\Omega\, D^{I\, ^\star}_{MK} (\Omega )\, D^{I'}_{M' K'} (\Omega )\,
D^{I''}_{M'' K''} (\Omega ).
\end{eqnarray}
In the case of $I+I'+I''$ being half-integer, the integration over the single
volume must be replaced by integration over the double volume~\cite{(Var88)}.
The integral in Eq.~(\ref{etilde2}) is equal~\cite{(Var88)} to
\begin{eqnarray}\label{integ}
\frac{2I+1}{8\pi^2}\int d\Omega\, D^{I\, ^\star}_{MK} (\Omega )\, D^{I'}_{M' K'} (\Omega )\,
D^{I''}_{M'' K''} (\Omega ) \nonumber \\= {\rm {\bf C}}^{IM}_{I''M''I'M'} {\rm {\bf C}}^{IK}_{I''K''I'K'},
\end{eqnarray}
where symbols ${\rm {\bf C}}$ stand for the Clebsch-Gordan coefficients. The integral has the same form both for
integer and half-integer angular momenta.

Inserting (\ref{integ}) to (\ref{etilde2}), and requesting that Eq.~(\ref{true}) holds at $n=1$,
gives rise to a set of linear equations for regularized matrix elements $\tilde{V}^{{\rm 2B}}_{I'M'K'}$:
\begin{equation}\label{set1}
V^{{\rm 2B,1}}_{IMK} = \sum_{I'M'K'} A^{IMK}_{I'M'K'} \tilde{V}^{{\rm 2B}}_{I'M'K'} ,
\end{equation}
where
\begin{equation}\label{acof}
A^{IMK}_{I'M'K'} = \sum_{I''M''K''} c_{I''M''K''}^{\rm{{\cal N}}}
{\rm {\bf C}}^{IM}_{I''M''I'M'} {\rm {\bf C}}^{IK}_{I''K''I'K'} .
\end{equation}
Matrix $A^{IMK}_{I'M'K'}$ is quadratic for even-even and odd-odd
nuclei and rectangular for odd-$A$ nuclei. The problem of finding the
regularized matrix elements within the LR scheme is thus reduced to
calculating auxiliary quantities (\ref{etilde}) for $n=1$ and then
solving a set of linear equations (\ref{set1}). In the HFODD solver,
the latter is handled by using the singular-value-decomposition (SVD)
technique.

We note here that the regularization procedure can be applied
separately to all terms of the EDF, that is, terms that correspond to
interactions can be treated within the standard AMP method, and only
those which do not, should be treated within the regularization
scheme.

The expansion of Slater determinant $|\Psi \rangle$ in terms of the AMP states
reads~\cite{(RS80)},
\begin{equation}
  |\Psi \rangle = \sum_{IK} |IKK\rangle = \sum_{IK} \hat{P}^I_{KK} |\Psi \rangle .
\end{equation}
In turn, the sum rule, which connects mean-field averages and projected matrix elements, has the form
\begin{equation}\label{sumr}
  \langle \Psi | \hat V_{\rm{2B}} |\Psi \rangle = \sum_{IK} \langle \Psi | \hat V_{\rm{2B}}
  \hat{P}^I_{KK} |\Psi \rangle  = \sum_{IK} V_{IKK}^{\rm{2B}}.
\end{equation}
The sum rule expresses the HF mean-field average value in terms of
the projected matrix elements, and thus constitutes a stringent test
of the performed AMP. On the one hand, when $\hat V_{\rm{2B}}$ is a {\it
true} interaction, the sum rule must be strictly obeyed. On the
other hand, for singular energy kernels, its violation gives a
numerical estimate of problems related to not using {\it true}
interactions. Similarly, the sum rule calculated for regularized
matrix elements,
\begin{equation}\label{sumreg}
  \langle \Psi | \hat V_{\rm{2B}} |\Psi \rangle = \sum_{IK} \tilde{V}_{IKK}^{\rm{2B}},
\end{equation}
tests the quality of the regularization procedure.
Note that sum rules must be obeyed separately for all terms in the
interaction, which allows for studying singularities of energy kernels
of separate terms.
In what follows, we show results obtained for the sum-rule residuals,
that is, for differences between right- and left-hand sides of
Eqs.~(\ref{sumr}) and~(\ref{sumreg}).
Apart from those, we also assess precision of the AMP by considering
energy $E_{I=0}$ of the lowest $I=0$ state.

More precisely, we focus our attention on investigating stability of
these two quantities in function of the highest angular momentum
${I_{\text{max}}}$ included in the calculations, and we present them
versus ${I_{\text{max}}}$. The same value of ${I_{\text{max}}}$ is
consistently used to define both summation ranges in
Eqs.~(\ref{set1}) and (\ref{acof}). Note that in Eq.~(\ref{acof}),
the range of summation should be higher than the natural cutoff
dictated by the highest meaningful AMP components in the mean-field
wave function, which are given by the values of amplitudes
$c_{IMK}^{\rm{{\cal N}}}$. With increasing values of
${I_{\text{max}}}$, the residuals of sum rules (\ref{sumr}) and
(\ref{sumreg}), should converge to zero.

All calculations were performed using the unrestricted-symmetry
solver HFODD~\cite{(Dob09d),(Sch12)}. We employed the Gauss-Chebyshev
quadratures to integrate over the $\alpha$ and $\gamma$ Euler angles
and the Gauss-Legendre quadrature to integrate over the $\beta$ Euler
angle. To achieve a sufficient accuracy,  for each Euler angle we
used a large number of mesh points equal $N_\alpha = N_\beta =
N_\gamma \equiv N = 50$.

The examples
presented below pertain to odd-odd nucleus $^{26}$Al, and to the so
called anti-aligned mean-field configuration, which is relevant in
the context of the superallowed Fermi $\beta$-decay~\cite{(Sat11)}.
The most demanding task was to calculate the auxiliary integrals
$V^{{\rm 2B,1}}_{IMK}$, Eq.~(\ref{etilde}). Since we were interested
in comparing the standard and regularized calculations, we decided
to use a relatively small configuration space, consisting of only $N_{\rm
shell}=6$ spherical harmonic-oscillator shells. Such small space
suppresses high angular-momentum components in the reference Slater
determinant. Unless explicitly stated, in all calculations, in both
direct and exchange channels the Coulomb interaction was treated
exactly.

\begin{figure}[htb]
\centering
\includegraphics[width=0.8\columnwidth]{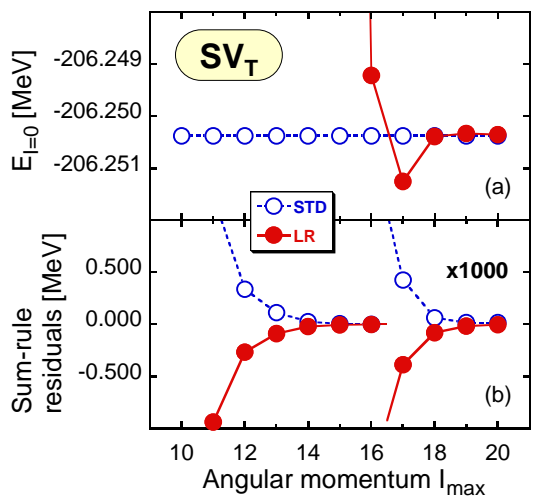}
\caption{(Color online) Convergence of the lowest $I$=0 energy (top) and Skyrme-energy
sum-rule residuals (bottom) in function of the
highest angular momentum ${I_{\text{max}}}$ included in the
calculations. Open and full circles represent results obtained using
the standard AMP method and our LR method, respectively. Calculations
were performed for the {\it true\/} Skyrme interaction SV$_{\rm T}$
(that is, with the tensor EDF terms included).}
\label{fig01}
\end{figure}

It is instructive to begin the discussion by showing results for
$E_{I=0}$ and Skyrme-energy sum-rule residuals obtained for the
SV$_{\rm T}$ Skyrme force. Such a calculation tests the numerical
implementation of the method, and can be regarded as a proof of
principle of the LR scheme. The reason is, as already mentioned, that
the SV$_{\rm T}$ is a {\it true\/} interaction and, therefore, both
standard AMP method and LR method should give exactly the same values
of both indicators. As can be seen in Fig.~\ref{fig01}, this is
indeed the case. It turns out that for ${I_{\text{max}}}\geq10$, the
standard AMP values of $E_{I=0}$ are perfectly stable (up to a
fraction of eV). However, the sum rule, which also tests the
convergence of higher angular momenta, reaches a similar level of
precision only above ${I_{\text{max}}}=20$. The LR values of $E_{I=0}$
converge only above ${I_{\text{max}}}=20$, which illustrates the fact
that in Eq.~(\ref{set1}), higher intermediate angular momenta must be
taken into account. Note, however, that the sum rules calculated
using both methods converge in a similar smooth way.

\begin{figure}[htb]
\centering
\includegraphics[width=0.8\columnwidth]{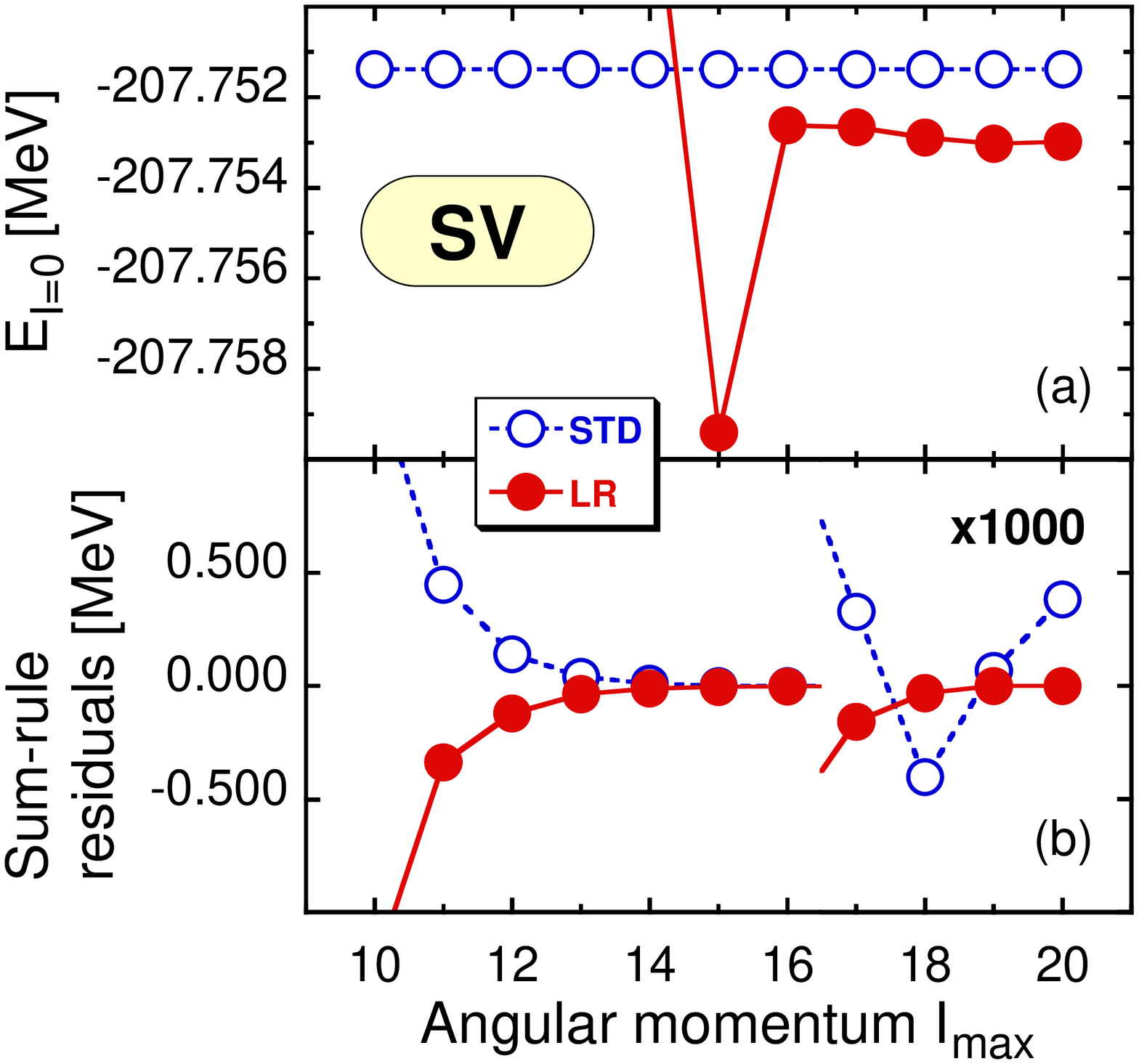}
\caption{(Color online) Same as in Fig.~\ref{fig01} but for the original Skyrme functional SV,
that is, with the tensor EDF terms neglected.
}
\label{fig02}
\end{figure}

The density-independent Skyrme parametrization SV in its original
formulation~\cite{(Bei75)}, that is, without the tensor EDF terms, no
longer corresponds to an interaction. Fig.~\ref{fig02} clearly shows
that even such a seemingly insignificant departure from the {\it
true\/} Hamiltonian is immediately detectable through the indicators
tested in our study. In the standard AMP, energy $E_{I=0}$ is again
perfectly stable over the entire range of studied values of
${I_{\text{max}}}$. However, such stability can be misleading,
because the LR value, which converges only at ${I_{\text{max}}}=20$,
differs by as much as 2\,keV.

Note that the singularity of energy kernels leaves its fingerprint in
the values of the standard-AMP sum-rule residuals. After an apparent
convergence (at the level of a few keV), which is visible below
${I_{\text{max}}}=16$, at the level of a few eV, this indicator, in
fact, does not converge to zero. On the other hand, the LR sum-rule
residuals smoothly converge to zero with high precision. An important
conclusion obtained here is that the stability of the ground-state
energy does not necessarily warrant that its value be free from
spurious effects.

\begin{figure}[htb]
\centering
\includegraphics[width=0.8\columnwidth]{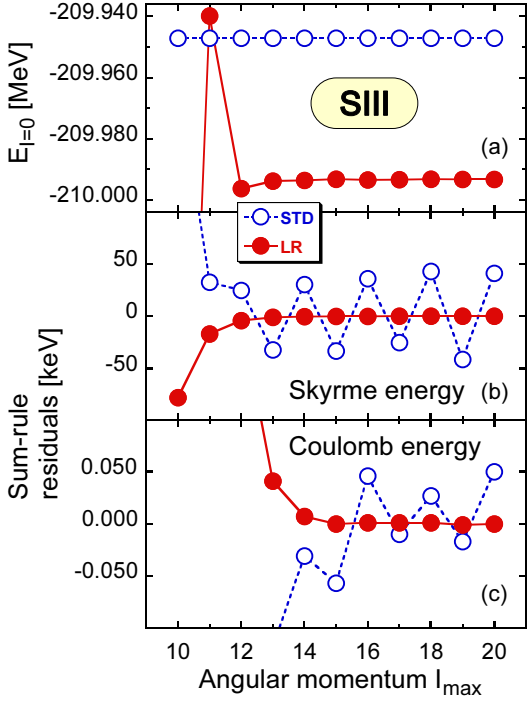}
\caption{Same as in Fig.~\ref{fig01} but for the original SIII Skyrme functional.
Panels (b) and (c) show sum-rule residuals of the Skyrme and Coulomb energies, respectively.
Slater approximation of the Coulomb exchange energy was used.
}
\label{fig03}
\end{figure}

For the density-dependent SIII functional~\cite{(Bei75)}, problems
encountered within the standard AMP are further magnified. In this
example, we performed calculations using the Slater
approximation~\cite{(Sla51)} of the Coulomb exchange energy.
Therefore, here we applied our LR method both to the Skyrme and
Coulomb parts of the functional. The results are depicted in
Fig.~\ref{fig03}, showing the energy $E_{I=0}$ (upper panel),
Skyrme-energy sum-rule residuals (middle panel), and Coulomb-energy
sum-rule residuals (lower panel). Similarly as in Fig.~\ref{fig02},
the standard AMP leads to misleadingly stable values of $E_{I=0}$; however, now
the corresponding sum rules turn out to be completely unstable. For
the Skyrme and Coulomb energies, they stagger around zero at the
level of 50\,keV and 50\,eV, respectively. In contrast, the LR method
perfectly stabilizes the sum rules, which smoothly converge to zero,
and leads to stable values of $E_{I=0}$. However, the LR $E_{I=0}$
energy is now shifted down by almost 50\,keV, as compared to the
standard AMP solution.

\begin{figure}[thb]
\centering
\includegraphics[width=0.8\columnwidth]{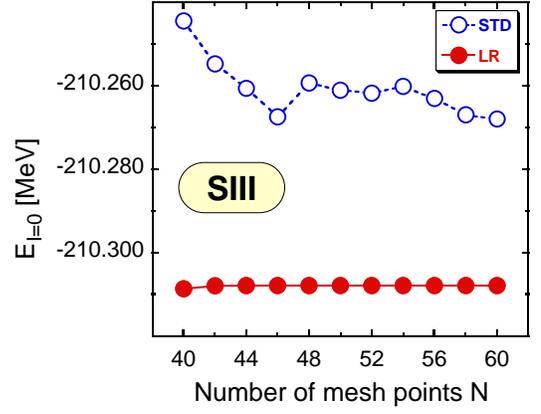}
\caption{(Color online) Energies $E_{I=0}$ calculated using the standard AMP (open circles) and
LR (full circles) methods, as functions of the number of mesh points $N$. Calculations were
performed for the original SIII Skyrme functional. The maximum angular momentum
of ${I_{\text{max}}}=20$ was used.
}
\label{fig04}
\end{figure}

Finally, let us point out yet another shortcoming of the standard AMP
approach. Fig.~\ref{fig04} shows results of the test of stability of
$E_{I=0}$ with respect to the number of mesh points $N$ used in
numerical integrations over the Euler angles. In this example,
calculations were performed for the SIII functional~\cite{(Bei75)} and
exact Coulomb exchange energy. It is clearly visible that the
standard-AMP values of $E_{I=0}$ vary strongly and quite erratically
with $N$. This is owing to the fact that the results do depend on
relative positions of mesh points with respect to singularities of
the energy kernel. In contrast, the LR results are perfectly stable.

\subsection{Quadratic regularization scheme}\label{QRS}

In the QR scheme, matrix element $V^{{\rm 2B}}_{IMK}$ (\ref{vimk}) is replaced by the auxiliary
quantity $V^{{\rm 2B,2}}_{IMK}$ defined in Eq.~(\ref{etilde}) for $n=2$.
Again we assume that matrix element $\langle \Psi |\hat V_{\rm{2B}} | \tilde{\Psi}\rangle$ is regularizable.
Inserting expansions (\ref{skex}) and (\ref{ovex}) into  (\ref{etilde}) gives
\begin{eqnarray}\label{etildeQ2}
V^{{\rm 2B,2}}_{IMK} =  \sum_{I_1 M_1 K_1} \tilde{V}_{I_1 M_1 K_1}^{\rm{2B}}
\sum_{I_2 M_2 K_2 } c_{I_2 M_2 K_2}^{\rm{{\cal N}}}
\sum_{I_3 M_3 K_3 } c_{I_3 M_3 K_3}^{\rm{{\cal N}}} \nonumber \\
\frac{2I+1}{8\pi^2} \int d\Omega\, D^{I\, ^\star}_{MK} (\Omega )\, D^{I_1}_{M_1 K_1} (\Omega )\,
D^{I_2}_{M_2 K_2} (\Omega )\, D^{I_3}_{M_3 K_3} (\Omega ). \nonumber \\
\end{eqnarray}
The integral can be calculated with the aid of the following Clebsch-Gordan series~\cite{(Var88)}:
\begin{eqnarray}\label{CGseries}
D^{J_1}_{M_1 K_1} (\Omega ) &\,& D^{J_2}_{M_2 K_2}  (\Omega ) = \nonumber \\
\sum_{J_3 = |J_1 -J_2|}^{J_1+J_2} \sum_{M_3 K_3} {\rm {\bf C}}^{J_3 M_3}_{J_1 M_1 J_2 M_2} &\,& D^{J_3}_{M_3 K_3}  (\Omega )
{\rm {\bf C}}^{J_3 K_3}_{J_1 K_1 J_2 K_2}.
\end{eqnarray}
This gives
\begin{eqnarray}\label{etildeQ3}
V^{{\rm 2B,2}}_{IMK} =  \sum_{I_1 M_1 K_1} \tilde{V}_{I_1 M_1 K_1}^{\rm{2B}}
\sum_{I_2 M_2 K_2 } c_{I_2 M_2 K_2}^{\rm{{\cal N}}}
\sum_{I_3 M_3 K_3 } c_{I_3 M_3 K_3}^{\rm{{\cal N}}} \nonumber \\
\sum_{I'' = |I_2 -I_3|}^{I_2+I_3} \sum_{M'' K''} {\rm {\bf C}}^{I'' M''}_{I_2 M_2 I_3 M_3}
{\rm {\bf C}}^{I'' K''}_{I_2 K_2 I_3 K_3} \nonumber \\
\frac{2I+1}{8\pi^2} \int d\Omega\, D^{I\, ^\star}_{MK} (\Omega )\, D^{I_1}_{M_1 K_1} (\Omega )\,
D^{I''}_{M'' K''} (\Omega ).\nonumber \\
\end{eqnarray}
Eventually, after using expression (\ref{integ}), we obtain
\begin{eqnarray}\label{etildeQ4}
V^{{\rm 2B,2}}_{IMK} =  \sum_{I_1 M_1 K_1} \left\{ \sum_{I_2 M_2 K_2} c_{I_2 M_2 K_2}^{\rm{{\cal N}}}
\sum_{I_3 M_3 K_3} c_{I_3 M_3 K_3}^{\rm{{\cal N}}} \right. \nonumber \\ \left.
\sum_{I'' = |I_2 -I_3|}^{I_2+I_3} \sum_{M'' K''} {\rm {\bf C}}^{I'' M''}_{I_2 M_2 I_3 M_3}
{\rm {\bf C}}^{I'' K''}_{I_2 K_2 I_3 K_3} {\rm {\bf C}}^{I M}_{I_1 M_1 I'' M''} \right. \nonumber \\
\left. {\rm {\bf C}}^{I K}_{I_1 K_1 I'' K''} \right\} \tilde{V}_{I_1 M_1 K_1}^{\rm{2B}} \equiv
\sum_{I_1 M_1 K_1} A^{IMK}_{I_1 M_1 K_1} \tilde{V}_{I_1 M_1 K_1}^{\rm{2B}} ,
\end{eqnarray}
with the following selection rules on intermediate summations:
\begin{eqnarray}
M_2+M_3 = M'' ,\quad  M_2+M_3 = M'', \nonumber \\
M_1+M'' = M   ,\quad  K_1+K'' = K .
\end{eqnarray}
From the practical point of view, it is important to notice that the intermediate summations
over $I'',M'',K''$ can be extended as,
\begin{equation}
\sum_{I'' = |I_2 -I_3|}^{I_2+I_3} \longrightarrow
\sum_{I'' = 0}^{2J},
\end{equation}
with the added terms being zero owing to properties of the Clebsch-Gordan coefficients.
This allows us to change in Eq.~(\ref{etildeQ4}) the order of summations, and to split the multi-dimensional
summations into two independent sets of sums, that is,
\begin{eqnarray}\label{aQ}
 \!\!\!\! A^{IMK}_{I_1 M_1 K_1} & = & \!\!\!\!
 \sum_{I'' M'' K''} X^{I''}_{M'' K''} {\rm {\bf C}}^{I M}_{I_1 M_1 I'' M''} {\rm {\bf C}}^{I K}_{I_1 K_1 I'' K''},
\end{eqnarray}
where
\begin{eqnarray}\label{xQ}
X^{I''}_{M'' K''} = \sum_{I_2 M_2 K_2} c_{I_2 M_2 K_2}^{\rm{{\cal N}}}
\sum_{I_3 M_3 K_3} c_{I_3 M_3 K_3}^{\rm{{\cal N}}} \nonumber \\
{\rm {\bf C}}^{I'' M''}_{I_2 M_2 I_3 M_3} {\rm {\bf C}}^{I'' K''}_{I_2 K_2 I_3 K_3}.
\end{eqnarray}
The trick used above facilitates numerical calculations. It should also be noted that the Clebsch-Gordan coefficients
impose selection rules that introduce further simplifications. For example, in Eq.~(\ref{aQ}), one has $M''=M-M_1$ and  $K''=K-K_1$, meaning
that corresponding summations are effectively one-dimensional.

In the applications of our QR scheme, we used a strategy similar to the one
used for the LR scheme, that is, energies $E_{I=0}$ and sum-rule residuals were investigated as a
functions of ${I_{\text{max}}}$, which is the maximum value of angular momentum admitted in
the summations in Eqs.~(\ref{aQ}) and (\ref{xQ}).

\begin{figure}[htb]
\centering
\includegraphics[width=0.8\columnwidth]{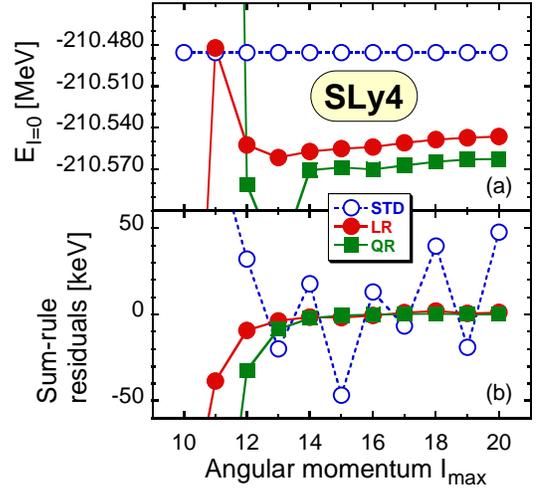}
\caption{(Color online) Same as in Fig.~\ref{fig01} but for the SLy4 Skyrme functional.
Full squares represent results obtained using our QR method.
}
\label{fig05}
\end{figure}

First, we applied the QR method to the SIII case, already analyzed in
Sec.~\ref{LRS} in the context of the LR method, see Fig.~\ref{fig03}.
In this case, both regularization schemes are fully equivalent and
give identical results. This not only tests the code but also speaks
in favor of the reliability of both schemes.

Next, we applied both regularization schemes to the case of the SLy4
functional~\cite{(Cha97)}, which features a fractional-power density
dependence with $\eta =1/6$. The results are presented in
Fig.~\ref{fig05}. In this case, the regularization is insufficient to
stabilize the energy of the lowest $I$=0 state. By removing
contributions coming from uncompensated poles in energy kernels, the
QR scheme lowers the energy of the $I=0$ state as compared to the LR
scheme. However, because of non-analyticity caused by the fractional
power, it is unable to fully stabilize the solution.

\section{Summary and perspectives}
\label{sum}

In this work, we proposed a method to regularize the two-body
off-diagonal MR DFT kernels and we presented the first application
thereof to a representative case of the angular-momentum projection.
The method is based on two general assumptions:
\begin{enumerate}
\item
First, it assumes that the MR EDF is regularizable, meaning in
practice that there exist a regularization scheme replacing the
kernel by its smooth counterpart that can be expanded in a set of
Wigner $D$-functions (\ref{skex}). No explicit knowledge of the
regularization scheme is required.

\item
Second, it assumes that the singularities, which appear in the
denominator of the two-body kernel, originate from vanishing overlap,
which is a result of using the GWT. This allows
us to identify terms proportional to $\langle \Psi
|\tilde\Psi\rangle^{-(1+\eta )}$, where $\eta$ comes from the direct
density dependence of the generator of two-body kernel, as
potentially the most dangerous ones.

\end{enumerate}
The essential advantage of the method is in avoiding the necessity
to explicitly remove the self-interactions~\cite{(Lac09),(Ben09)}.
Instead, the proposed method relies on computing a set of auxiliary
integrals of non-singular kernels obtained by multiplying the
original ones with an appropriately chosen power of the overlap.
Provided that the GWT is the only source of singularities, the
auxiliary integrals turn out to be linear functions of the true
regularized matrix elements, and the entire problem can be reduced to
an algebraic task of solving a set of linear equation. The method
also has a certain internal flexibility, namely, it can be applied to
selected parts of the EDF only. This feature is important in cases
when troublesome parts of the EDF can be isolated. The example is the
Coulomb exchange interaction treated within the Slater approximation.

A certain disadvantage of the method is the fact that, to achieve a
desired accuracy, the auxiliary integrals require more quadrature
nodes. In addition, one has to invert or perform the SVD
decomposition of the auxiliary matrices $A$, see Eqs.~(\ref{acof})
and (\ref{aQ})), which can be a potential source of numerical
inaccuracies. Both problems become more acute with increasing power
of the overlap multiplying the kernel. Nevertheless, in this
exploratory study we have demonstrated that both the LR and QR
schemes can be realized. Problems encountered for the calculations
involving the SLy4 functional reflect the inadequacy of the scheme in
applications to fractional-power density-dependent terms, which lead
to non-analytic dependence of kernels on the Euler angles.

Finally, an interesting feature of our calculations is that they
reveal the fact of how much the effects of using EDFs that are not
derived from {\it true\/} interactions can be inconspicuous. Our
calculations show that even the slightest departure from the {\it
true\/} interaction is immediately detectable through the sum rules,
but it can be completely invisible when looking at energies of
low-lying states, which are often perfectly stable.

\begin{acknowledgments}

This work was supported in part by the Polish National Science Centre (NCN) under Contract No. 2012/07/B/ST2/03907,
by the THEXO JRA within the EU-FP7-IA project ENSAR (No.\ 262010), by the
ERANET-NuPNET grant SARFEN of the Polish National Centre for Research and
Development, and by the Academy of Finland and University of Jyv\"askyl\"a within the FIDIPRO programme.

\end{acknowledgments}


%

\end{document}